# Decision Support for Increasing the Efficiency of Crowdsourced Software Development

Muhammad Rezaul Karim
University of Calgary
2500 University Drive NW
Calgary, Alberta T2N 1N4
+1 (403) 220 7692
mrkarim@ucalgary.ca

David Messinger
Topcoder
425 Market Street
San Francisco, 94105, CA
USA
+1 (978) 590-3344
dmessinger@topcoder.com

Ye Yang
Stevens Inst. of Technology
1 Castle Point Ter
Hoboken, NJ 07030, USA
+1(201)216-8560
ye.yang@stevens.edu

Guenther Ruhe
University of Calgary
2500 University Drive NW
Calgary, Alberta T2N 1N4
+1 (403) 220 7692
ruhe@ucalgary.ca

## ABSTRACT

Crowdsourced software development (CSD) offers a series of specified tasks to a large crowd of trustworthy software workers. Topcoder is a leading platform to manage the whole process of CSD. While increasingly accepted as a realistic option for software development, preliminary analysis on Topcoder's software crowd worker behaviors reveals an alarming task-quitting rate of 82.9%. In addition, a substantial number of tasks do not receive any successful submission.

In this paper, we report about a methodology to improve the efficiency of CSD. We apply massive data analytics and machine leaning to (i) perform comparative analysis on alternative technique analysis to predict likelihood of winners and quitters for each task, (ii) significantly reduce the amount of non-succeeding development effort in registered but inappropriate tasks, (iii) identify and rank the most qualified registered workers for each task, and (iv) provide reliable prediction of tasks risky to get any successful submission.

Our results and analysis show that Random Forest (RF) based predictive technique performs best among the alternative techniques studied. Applying RF, the tasks recommended to workers can reduce the amount of non-succeeding development effort to a great extent. On average, over a period of 30 days, the savings are 3.5 and 4.6 person-days per registered tasks for experienced resp. unexperienced workers. For the task-related recommendations of workers, we can accurately recommend at least 1 actual winner in the top ranked workers, particularly 94.07% of the time among the top-2 recommended workers for each task. Finally, we can predict, with more than 80% F-measure, the tasks likely not getting any submission, thus triggering timely corrective actions from CSD platforms or task requesters.

## 1. INTRODUCTION AND BACKGROUND

The most expensive part of software development is people. Even further, the most valuable asset of a company is its human resource. Treating them accordingly and organizing their work in an efficient manner is critical for project success. Crowdsourced software development (CSD) is directed towards higher efficiency, leveraging a large crowd of trustworthy software workers who are registering and submitting for their interested tasks in exchange of financial gains [3]. A general CSD process starts with task requesting companies distributing tasks with prizes online, and then crowd software workers browsing and registering to work on selected tasks, and submitting work products once completion. Crowd submissions will be evaluated by experts and experienced developers, through a peer review process, to check the code quality and/or document quality [1, 2]. The number of submissions and their evaluated scores reflect the level of success in task satisfaction or completion [3].

As one of the most successful CSD platforms, Topcoder has over 1 million registered workers from over 190 countries, averagely 80K logins every 90 days, 7K challenges hosted per year and $80M in challenges payouts. The size of crowd workers is almost 5 times more engineers than Microsoft, Facebook, and Twitter combined. However, utilizing unknown, external developers incurs new issues related to worker identification and trust management. For example, an analysis on Topcoder data from 2014-2015 shows an 82.9% of worker quitting rate, on average 55.8% submission not passing review, and a task cancellation rate of 15.7% [3]. In his keynote at the 3rd Workshop on Crowdsourcing in Software Engineering, Messinger recognized "trust and transparency" as one of three key elements of good CSD [4]. Accurate and timely analytics to support trust and transparency is critical for measuring and predicting worker reliability, process stability, and products quality in CSD context.

To that end, existing studies have focused on decision support for software crowdsourcing market. Among them, most focused on supporting decision making from the perspectives of task requesters or crowdsourcing platforms. These studies include task pricing [5], developer recommendations [6], and understanding worker behaviors [7, 8]. In our previous paper "Who Should Take This Task? – Dynamic Decision Support for Crowd Workers" [3], we proposed an analytical framework called *DCW-DS* and reported empirical results to predict the success of winners and quitters in CSD. More specifically, we provided recommendations to workers which task they should follow or register and which ones better not. For workers just following one of the top three task recommendations, we have shown that the average quitting rate goes down below 6%.

In this paper, driven by top business concerns in improving CSD efficiency according to Topcoder, we extend DCW-DS and conduct an industrial evaluation in four main directions:

- Comparative analysis of alternative classification techniques to justify the selection of Random Forest in DCW-DS.
- Taking the industrial perspective of Topcoder, we study the potential effort savings that would result from applying the recommendations given.
- Comparison of the actual winner scores with the scores of the ones recommended.
- Prediction of task cancellation and its potential effort savings.

One of the main strengths of this extension work is that on each day the output of a single prediction model is applied to for several purposes ranging from decision support for workers and task requesters. In more detail, we provide ranking of the tasks per worker, ranking of the workers per task, and predicting cancellation of tasks with all the same basic methodology, which is more efficient than building a separate prediction model for each purpose.

Section 2 gives the formulation of the research questions. Our predictive methodology is presented as Section 3, with presentation and discussion of empirical results being the focus of Section 4. Finally, discussions and conclusions are given in Section 5 and 6.

## 2. RESEARCH QUESTIONS

With the given industrial context and background, we have derived four research questions. All the RQ's are evaluated in the context of real-world data coming from Topcoder development projects. In what follows, we provide the four questions and their justification taken from an industrial application perspective.

**RQ1:** How does the performance of the previously selected RF algorithm compare with other learners such as SVM, NB, DT for predicting likelihood of winner and quitter workers?

**Why?** The first RQ is to justify the selected methodology of applying ensemble classifiers based on the idea of Random Forest (ensemble of decision tree analysis with randomized samples and attributes) [30]. For further details on the modeling and attributes used for performing the machine learning classification, see [3].

**RQ2:** For various tasks, how does the effort following RF recommendations workers could have saved compared with the actual total effort spent?

**Why?** This RQ is designed to measure the amount of effort saved from not performing non-succeeding development effort when workers would apply the generated recommendations.

**RQ3:** How does the avg. submission score of recommended most qualified registered workers compare with the avg. submission scores of actual winners for each task?

**Why?** This RQ is designed to measure the quality of the recommended most qualified registered workers for each task.

**RQ4:** Can DCW-DS be used in the prediction of task cancellation?

**Why?** This RQ is designed to measure how effectively the recommendation approach can detect the tasks to be cancelled due to insufficient number of submissions. In case successful this would help to adjust incentives to improve success rate.

## 3. METHODOLOGY

In what follows, we give a brief overview of the methodology applied. What we will see is how the general idea of DCW-DS with RF (Section 3.1) and its proposed extension can be used for the three subsequent research questions. This is described in Sections 3.2 to 3.4, respectively.

### 3.1 Overview

In our former work [3], we proposed DCW-DS where we built daily prediction models using Random Forest (RF) [11]. RF is an ensemble of classifiers and has outperformed other classifiers in other applications [11]. The real-world data used for the empirical study was taken from Topcoder development projects over a period between January 22, 2014 and March 9, 2015. From applying DCW-DS for a period of consecutive days, the output of the RF model was used for ranking the tasks per worker based on their suitability.

In DCW-DS, before building a model for a specified day, for each sample a collection of static and dynamic features were extracted. For details of the features we refer to [3] where it was shown that dynamic features substantially improve the quality of the classifications. Dynamic features were derived for a historical data of past T days. In this paper, the value of T was taken to be 90 days. This parameter setting was found empirically to be a good balance between information richness and information up-to-date-ness.

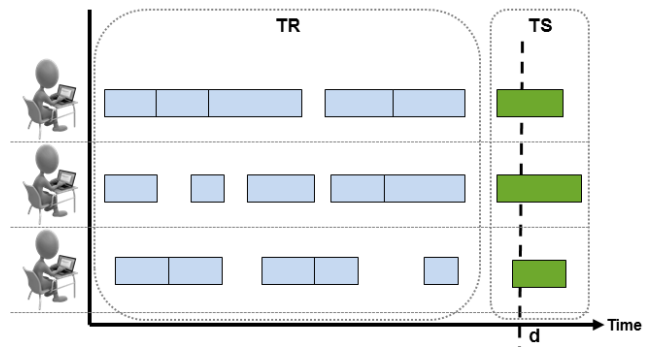

**Figure 1. Illustration of selecting training set TR and testing set TS with tasks performed over time by three workers.**

Before building and evaluating a model on a particular day $d$, we create two sets of data samples called *Training Set* (TR) and *Testing Set* (TS). The composition of the sets is illustrated in Figure 1. Each sample in the set TR represents information for a developer-task pair where the developer actually registered for a task and the task submission deadline was earlier than the current date $d$.

Similarly, each sample of TS contains same kind of information for all the tasks with submission deadline beyond date $d$ and task registration open date before day $d$ or on $d$. All samples in TS and TR are labeled as either *winner, quitter, or submitter* based on actual competition results. More specifically, these labels refer to having submitted and having been rewarded, having registered for a task but failed to submitting, or having submitted a task but not having been rewarded. Once training and test samples were created, we predicted the label for each sample in the test set TS and matched with the actual label. The winner and quitter are the most important labels from decision support perspective.

The WEKA machine and data mining library [12] was used for building and evaluating the models. The predictive modeling experiments conducted were performed with 124 features (excluding class variable), including 14 dynamic features and 110 static features [3]. Among the static features, 107 binary features encoded the required technologies (e.g., css, html5, Java) of the task. The rest of the static features were task duration, task total prize and overall submission rate of the worker. The whole datasets used in this study, including task attributes, worker attributes, and extracted features, is posted in [9].

In this work, as an extension of DCW-DS, we demonstrate the superiority of RF against other machine learners, we perform ranking of workers per task (Section 3.3) based on their potential success chance (winning and submission chance). On top of that, we perform prediction for tasks which likely would require cancellation (Section 3.4) due to zero qualified submissions.

## 3.2 Ranking Likely-Winning Tasks for Workers

For each worker from the test set, we rank the most relevant tasks. To come up with the ranking, we first identify the test samples belonging to each worker. Then we rank the identified samples in descending order of the workers' winner label probability score and put them in a list.

We discard the samples with low winner probability score (i.e., less than 0.33, or the winner probability score less than the submitter probability score). The probability score of 0.33 is chosen as threshold value as we have three classes (i.e. 1.00 divided by 3). Next, we sort the same identified samples in descending order based on their submitter probability score. In this case, we filter samples based on low submitter probability score (i.e., less than 0.33, or the submitter probability score less than the winner probability score). Then append the remaining samples in the tail of the previously constructed list if not already added.

The constructed lists contained ranked tasks for each worker with tasks with high winning chance followed by high submission chance. For some workers, especially with workers with no winning history, the constructed list contains ranked tasks with high submission chance only. Our hypothesis is that the application of this kind of task ranking of workers can reduce task quitting rate.

## 3.3 Ranking of Registered Workers for Tasks

At each day $d$, we first identify the registered workers for each task in the test set. Second, for each task, rank the workers in descending order based on their winning probability score in this task. Third, apply same kind of winner and submitter score based filtering like Section 3.2. Fourth, sort the same workers in descending order of their submitter label probability score and applied same kind of winner and submitter score based filtering as described in Section 3.2. Finally, append the remaining workers in the tail of the previously constructed list if already not added. Unlike ranking all potential workers [6] for a task, in this work we rank only the registered workers of a task on a daily basis.

## 3.4 Prediction of Task Cancellation

On each day $d$, first, we identify the tasks in the test set with duration greater than or equal to three days. Second, for each task in the test set, predict each registered worker in the test set as winner, quitter or submitter. Third, mark each task as 'Potentially Cancelled' for that day if the total number of recommended workers is zero on that day (section 3.3). Finally, predict a task as 'Cancelled' if the task has been marked as 'Potentially Cancelled' at least for the last $N-1$ days including the day $d$, where N is defined as three days.

## 3.5 Metrics for Performance Evaluation

For the goal of evaluating the quality of our predictions, we have defined five metrics to evaluate the accuracy of predictions:

**Definition 1**: *Precision* describes the percentage of samples of correctly predicted quitter (or winner or submitter).

**Definition 2**: *Recall* describes the percentage of samples of the quitter (or winner or submitter) class in the predicted results, out of all the samples that are quitter (or winner or submitter, respectively).

**Definition 3**: *F-measure* is the harmonic mean of precision and recall and combines these two measures into one.

**Definition 4**: *Recall@K* is the average of the probability of finding at least one of the workers out of the Top K recommended workers for all tasks.

**Definition 5**: *Score gap* is the average difference (gap) between the average review scores of actual winners and average review scores of the top two recommended workers computed for all tasks. The *Score gap* is always positive (zero or more) as the winners always have better review scores than the other workers. Score gap is zero when our approach recommends the actual winners in the top two positions for all tasks.

## 4. ANALYTICS RESULTS

In this section, we report the results from answering the four stated research questions.

### 4.1 RQ1: Comparative Analysis of Algorithms

Which algorithm is best for making our predictive analysis? We compared the performance of four established machine learning algorithms being Random Forest (RF), Support Vector Machines (SVM), Naïve Bayes (NB) and Decision Trees (DT). While more techniques exist, the emphasis later on was on having one proven very good technique and applying it to the different questions studied. To make the comparison meaningful, we identified the best parameter settings for each algorithm in terms of the F-measure values as well as the area under the ROC curve (AUC) [10] taken over 30-day period.

When applying RF, we varied the number of trees and the number of features used for classification. We tried five values (being 10, 25, 50, 75, 100) for the number of trees parameter and four values (10, 30, 50, 75) for the number of features parameter. In total, 20 parameter configurations were analyzed. As a result, we did not observe any statistically significant difference for the quitter class as well as winner class predictions in terms of F-measure value. No difference was also observed in terms of AUC. However, we achieved better results (average F-measure value better by 1% to 2%) for the winner class when the number of features was set to at least 50. Considering this, for our further experiments, we set the number of features to 50, while the number of trees parameter was specified to be 100.

For the SVM algorithm with polynomial kernel, we tried six different values for the complexity parameter: 0.01, 0.1, 0.25, 0.5, 0.75, 0.99 and three different values for the exponent parameter: 1, 5 and, 10. In total, we had 18 different configurations. For the decision tree, the values for the confidence factor and minimum number instances per leaf was varied. For the confidence factor we evaluated six different values: 0.01, 0.1, 0.25, 0.5, 0.75, 0.99. For the minimum number of instances per leaf parameter, we tried three different values: 2, 5, and 10. For the Naïve Bayes parameter, no tuning parameter was available from WEKA.

For SVM, when the complexity parameter was set to 0.75, we observed statistically significantly better performance than with other complexity parameter values, regardless of the chosen exponent parameter values. For the further experiments with SVM, we set the value of the complexity parameter to 0.75, while the exponent parameter value was arbitrarily set to 1.

For DT, we also did not observe any statistically significant difference in terms of F-measure values for the winner and quitter class as well as in terms of AUC. Like RF, the best parameters for DT were chosen considering the better average F-measure values over 30 days. For DT, for further experiments, the confidence factor parameter was set to 0.25, while the minimum number instances per leaf was set to 5.

For the comparison between the different algorithms, we selected three 30 days long time periods starting at 1 April 2014, 1 September 2014, and 1 January 2015. In Figure 2, for each algorithm, we report the average of the various performance metrics taken over all 90-days data. Next, we summarize our main findings from applying the above parameter settings:

**Finding 1.1 (Quitter classification):** RF based predictive model achieved average precision, recall and F-measure value of more than 98% for the quitter class.

**Finding 1.2 (Winner classification):** For the winner class, RF achieved 84% average precision, 87% average recall, and 85% average F-measure.

**Finding 1.3 (Size of training set):** Even though each of these 90 days have very different number of training samples (between 4743 to 27210 samples), performance did not have much impact. The average precision of 84%, average recall of 87% and average F-measure value of 85% for the winner class indicate that RF based model can successfully be built from a set of few thousands of training samples.

**Finding 1.4 (Comparison between algorithms):** Besides RF, DT based predictive models achieved next best results. The results of the comparison between algorithms are summarized in Figure 2.

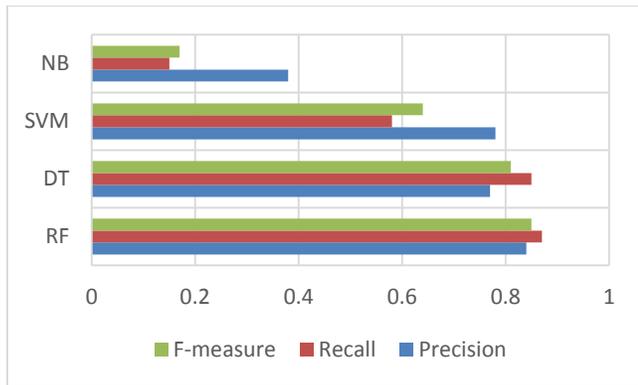

**Figure 2. Performance evaluation on 90 days (30 consecutive days starting at 01 April 2014, 01 September 2014 and 01 January 2015 ) in terms of average Precision (P), average Recall (R) and average F-Measure (F).**

With our *Vargha-Delaney* effect size comparison, we noticed that RF outperforms other algorithms in terms of precision, recall and F-measure with high probability. So, in our further experiments, we applied the RF algorithm with the best obtained parameter setting.

**Finding 1.5 (Analysis for specific types of tasks):** When we look at the results for three important task types only: *Assembly*, *Code*, and *UI Prototype*, in some cases we noticed better results. For *Assembly* type, RF achieved 85% average precision, 96% average recall and 90% average F-measure value for the winner class. For *UI Prototype* type, accuracy was 82%, 91% and 85% respectively for the same class. Finally, we observed 77% average precision, 89% average recall and 82% average F-measure value for the winner class for *Code* type.

## 4.2 RQ2: Savings in Development Effort

To answer RQ2, we measured the amount of effort that could be saved over 30 consecutive days. The amount of effort is averaged over all distinct task-worker pairs for this period of time. For each worker registered on a task, effort saving is measured as the number of days the prediction algorithm correctly predicted the worker as quitter for that task. For that, our assumption is that once an actual quitter registered on a task, he/she continued working on that task and subsequently failed to submit and became quitter. As it is not possible to exactly figure out how many hours a worker spends on a task per day basis (when the worker concurrently working on multiple tasks) and how many hours s/he spends in total on all tasks per day basis, we report our effort savings in person-days, where each person-day is equal to one calendar-day.

In Table 1, we report savings results for three periods of 30 consecutive days. Results were compared between for two different classes of workers: unexperienced (having submitted not more than 10 tasks successfully in the last 90 days) vs experienced workers.

**Table 1. Effort savings (person days) over 30 consecutive days. "Total" ("Avg.") describes the total (resp. average) effort savings over 30 days and all worker-tasks pairs.**

| Time period Starting at | Experienced | | | Unexperienced | | |
|---|---|---|---|---|---|---|
| | Total | Avg. | #Pairs | Total | Avg. | #Pairs |
| 1 Apr. 2014 | 59 | 3.69 | 59 | 9724 | 4.57 | 2127 |
| 1 Sep. 2014 | 167 | 3.63 | 46 | 11829 | 4.59 | 2573 |
| 1 Jan. 2015 | 166 | 3.39 | 49 | 10322 | 4.70 | 2197 |

**Finding 2.1 (Effort savings across tasks):** For all unexperienced workers, on average, around 4.57 to 4.70 person days were saved. If we look at the results for the experienced workers, we notice that total effort savings were a bit lower but still significant.

**Finding 2.2 (Task-specific effort savings):** There were no significant differences between the three most important task types (*Assembly*, *Code* and *UI Prototype*).

**Finding 2.3 (Effort savings without assembly type):** Currently, Topcoder allows to unregister only for *Assembly* task type, within 48 hours of registration. Due to lack of data availability for the specific assembly type, in Table 1, we used the above assumptions for all task types. However, when we look at the results for all task types except assembly type, the average savings for experienced and unexperienced workers slightly drops (around 4.03 on average for unexperienced workers, 3.17 on average for experienced workers). We also noticed that the number of task-pairs per 30-day period reduces to almost half for the unexperienced category (1087 task-worker pairs for 1 September 2014 period, 1038 task-worker pairs for 1 January 2015 period).

## 4.3 RQ3: Quality of Recommended Workers per Task

To answer this RQ, we ranked the recommended workers per task on a daily basis and measured the average difference (gap) between the average review scores of actual winners vs. the ones of the top two recommended workers per task. In addition, we measured the probability of finding at least one of the winners in the top 1 and top 2 recommended workers (recall@1 resp. recall@2).

To illustrate the whole scoring and recommendation process, we provide an **Illustrative example.** For that, we consider two tasks (30047945 and 30048207) from the test set TS of Jan 1st, 2015. For the first task, our approach recommended two workers: *GreatKevin* and *suno1234*. After few days, after the task submission deadline, review was done by Topcoder, *GreatKevin* had final review score of 98.75, while *suno1234* achieved 95.63. These recommended workers also became winners. In this case, as average score of our top two recommended workers (98.75 + 95.63)/2 is same as the average score of actual winners, so the score gap is zero.

For the other task, three workers *albertwang*, *seriyvolk83*, and *mohamede1945* were recommended by our approach in this order of preference. This time, *mohamede1945* and *seriyvolk83* were selected as the ultimate winners. That means, our approach ranked one of the winners (*mohamede1945*) as number 3. So, in this case, the gap between the average score of two actual winners (99.82 + 98.78)/2 = 99.30 and the average score (95.01 + 98.78)/2 = 96.90 of our top two recommended workers is 99.30 - 96.90 = 2.4.

After the illustrative example, we now continue with the two main findings related to RQ3:

**Finding 3.1 (Quality of recommendations across all task types):** Our prediction generates very high quality of recommendations for tasks both in terms of average score gap (being 2.41%), Recall@1 (87.74%) and Recall@2 (94.07).

On average, over 90 days, the score gap is 2.41 with low variance. On average, 87.85% of the time we can find at least one of the winners in the top ranked recommended worker and 94.07 % of the time among the top two recommended workers. In both cases, we had very low variation across different days. In Figure 3 and 4, we report the day wise performance (daily average taken over all tasks in the relevant test set) over 90 different days.

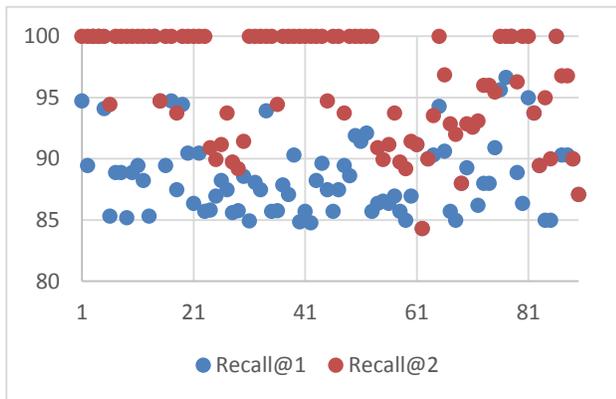

**Figure 3. Recall@k performance for 90 different days.**

**Finding 3.2 (Quality of recommendations across for specific types):** For *Assembly* type, recall@1 and recall@2 was 93.58 and 99.02 respectively. For the code type, recall@1 and recall@2 was 87.27 and 96.28 respectively. For the *UI prototype* type, the value for recall@1 was 90.1%, while recall@2 was 97.5%. For all three types (i.e. *Assembly, Code and UI Prototype*), the average score gap was 1.00%, 2.61% and 1.19% respectively.

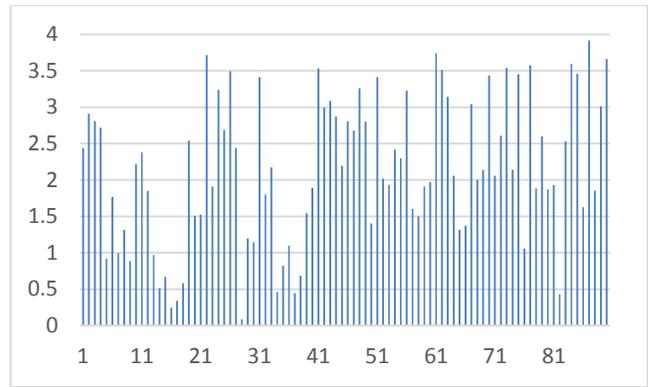

**Figure 4. Average score gap of 90 different days.**

## 4.4 RQ4: Prediction of Task Cancellation

For this RQ, we have taken the average of the precision, recall and F-measure values for two 30-day periods. In Table 2, we report the results for those time periods.

**Finding 4.1 (Task cancellation prediction accuracy):** Using our framework, we can predict cancellation of tasks with high precision, recall and F-measure. In both cases, the precision was at least 85%.

**Table 2. Task prediction performance**

| Time Period | Precision | Recall | F-measure |
|---|---|---|---|
| 1 Sep 2014 | .85 | .80 | .82 |
| 1 Jan 2015 | .87 | .77 | .81 |

**Finding 4.2 (Effort savings from task cancellation prediction):** On average 54.81% of the actual task duration were saved for the 66 tasks found in the test sets of the time periods in Table 2 (always deducting three days used for monitoring). The boxplot in Figure 5 shows the distribution of the percentage of savings in terms of task duration of different tasks. The significant time savings clearly indicates that monitoring each tasks for three consecutive days for zero recommended workers are sufficient to predict them as *cancelled*.

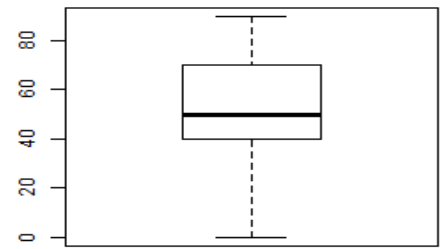

**Figure 5. Percentage of savings in terms of task duration**

In our computation, task duration is defined as the task submission end date minus task registration start date. We also assume here is that the tasks are cancelled once recommended. For our 66 tasks, minimum, maximum and average task duration was 4, 30 and 10.78 calendar days, respectively. The higher the actual duration, the higher was the savings (90% savings was achieved for tasks with 30 days duration, 60% savings for tasks with 7 days duration).

## 5. DISCUSSION

How valid are the conclusions taken and how practical are the results achieved from Topcoder's perspective? We discuss these two questions in Sections 5.1 and 5.2, respectively.

### 5.1 Threats to Validity

Parameter tuning of prediction models can impose construct validity threat for the conducted analysis. To reduce this threat related to construct validity, we performed parameter tuning and chose the best configuration for each algorithm using F-measure value and AUC as selection criteria. For each algorithm, we tried different parameter configurations. The data sets used in the analysis can impose external validity threat. To minimize this threat, we used a data set containing data from diverse number of projects from different time frame, with thousands of tasks of different task types, platform and required technologies. Selection of consecutive days for performance evaluation can also be a threat to conclusion validity. To minimize this threat, we used 30 consecutive days training and test data from three different time frames (90 days in total). Each time frame also contains varying number of training and test samples.

### 5.2 Applicability of Results to Topcoder

Efficiency of CSD processes is a key business driver for Topcoder. Results reported in this study indicate the following main benefits for the company for improving their CSD efficiency:

**1) Informed task/worker selection.** On the one hand, providing crowd workers with information on their competitive status allows them to judge and decide earlier which tasks are most promising to pursue. The results of RQ 2 indicate that there is a strong potential to focus on most tasks being closest to the workers track record and expertise. For potential submitters, being smarter in task selection saves significant amount of development effort spent but wasted; for potential quitters, being smarter in task selection prevents decreasing of worker reliability score due to registering and later quitting. On the other hand, task requesters and CSD platform providers are also supported with information regarding the most appropriate workers (results of RQ3), which can be used to more proactively invite and interact with targeted workers.

**2) Quality of task deliverables.** RQ3 results indicate that our recommendations are strong in terms of covering the actual winner either as the top ranked or the second ranked worker. This provides another argument on the quality of the classifications made. For each task, the ranking of the workers generated in this approach first takes the winning chance of workers into account. The workers who have high track record of winning and potentially submitting high quality solutions on a task are ranked higher. Task wise worker ranking can be used by Topcoder for prioritizing the potential top K (e.g., K = 5) submissions that would only be reviewed (if they want) to reduce huge time commitment involved in the review process of a task. In addition, this type of ranking can help to identify the top K submissions where most qualified reviewers should be assigned to. For higher efficiency, expert reviewers should review the high quality submissions preferably.

**3) Early warning of task cancellations.** RQ4 addresses the CSD monitoring challenge in how to find and subsequently treat tasks not receiving any successful submission. Whatever the reasons are for this situation, the number of these tasks is non-trivial and requires an early response from the provider. The capability to predict this type of vulnerable tasks is helpful for to intervene and mitigate task failure risk in a timely manner, through either re-scoping, further refinement, or adjusting task pricing.

## 6. CONCLUSIONS

While crowdsourced development is more and more accepted as a competitive alternative to other development paradigms [13], we could provide evidence that machine learning predictions based on Random Forest classification can be supportive to make the process even more efficient. From applying the same methodology, a variety of questions could be answered. Elimination of wasted time for developers unlikely to succeed and early responses to tasks likely to not receiving qualified submissions are of immediate benefit to increase the transparency and trust across Topcoder community. Although all analysis was done following the Topcoder processes and data, the approach can be adopted to other providers as well. Follow-up investigations are needed and planned to compare the actual savings with the ones outlined as well as looking at other parts of the overall process.

## ACKNOWLEDGEMENTS

This research was partially supported by the Natural Sciences and Engineering Research Council of Canada, NSERC Discovery Grant 250343-12. One of the authors was supported by a grant from Alberta Innovates Technology Future.